\documentclass[longauth]{aa}

\usepackage{graphicx}
\usepackage{multirow}
\usepackage{tabularx}
\usepackage{longtable,tabu}
\usepackage{lscape}
\usepackage{subfigure}
\usepackage{natbib}

\def\sun{\odot}
\bibpunct{(}{)}{;}{a}{}{,}
\usepackage{multirow}

\usepackage{amssymb}
\usepackage{pifont}
\newcommand{\cmark}{\ding{51}}%
\newcommand{\xmark}{\ding{55}}%
\usepackage{url}
\usepackage{threeparttable}
\usepackage{lipsum,multicol}

\newcommand\methane{\ifmmode{{\rm CH}_{4}}\else{CH$_{4}$}\fi}
\newcommand\water{\ifmmode{{\rm H}_{2}{\rm O}}\else{H$_{2}$O}\fi}
\newcommand\carbdiox{\ifmmode{{\rm CO}_{2}}\else{CO$_{2}$}\fi}
\newcommand\ammonia{\ifmmode{{\rm NH}_{3}}\else{NH$_{3}$}\fi}
\newcommand\acetylene{\ifmmode{{\rm C}_{2}{\rm H}_{2}}
	\else{C$_{2}$H$_{2}$}\fi}

\usepackage{verbatim}

\begin{document}
	
	\title{The GAPS Programme at TNG}
	\subtitle{XXXVIII.Five molecules in the atmosphere of the warm giant planet WASP-69b detected at high spectral resolution\thanks{Based on observations made with the Italian Telescopio Nazionale Galileo (TNG) operated on the island of La Palma by the Fundacion Galileo Galilei of the INAF at the Spanish Observatorio Roque de los Muchachos of the IAC in the frame of the program Global Architecture of the Planetary Systems (GAPS).} }
	
	\titlerunning{Five molecules in the atmosphere of the warm giant planet WASP-69b}
	
	\authorrunning{Guilluy et al.}
	
	\author{G. Guilluy\inst{1,2} \and
		P. Giacobbe\inst{1} \and
		I. Carleo\inst{3,4} \and
		P. E. Cubillos\inst{1,5} \and
		A. Sozzetti\inst{1} \and
		A. S. Bonomo\inst{1} \and
		M. Brogi\inst{1,6,7} \and
		S. Gandhi\inst{6,7,8} \and
		L. Fossati\inst{5} \and
		V. Nascimbeni\inst{4} \and
		D. Turrini\inst{1} \and
		E. Schisano\inst{9} \and
		F. Borsa\inst{10} \and
		A. F. Lanza\inst{11} \and
		L. Mancini\inst{12,13,1} \and
		A. Maggio\inst{14} \and
		L. Malavolta\inst{4,15} \and
		G. Micela\inst{15} \and
		L. Pino\inst{16} \and
		M. Rainer\inst{10} \and
		A. Bignamini\inst{17} \and
		R. Claudi\inst{4} \and
		R. Cosentino\inst{18} \and
		E. Covino\inst{19} \and
		S. Desidera\inst{4} \and
		A. Fiorenzano\inst{18} \and
		A. Harutyunyan\inst{18} \and
		V. Lorenzi\inst{17} \and
		C. Knapic\inst{17} \and
		E. Molinari\inst{20} \and
		E. Pacetti \inst{9,21} \and
		I. Pagano\inst{14} \and
		M. Pedani\inst{18} \and
		G. Piotto\inst{4} \and
		E. Poretti\inst{9,18}
	}
	
	\institute{
		INAF -- Osservatorio Astrofisico di Torino, Via Osservatorio 20, 10025, Pino Torinese, Italy
		\and
		Observatoire Astronomique de l'Universit\'e de Gen\`eve, Chemin Pegasi 51b, 1290, Versoix, Switzerland
		\and
		Astronomy Department, Indiana University, Bloomington, IN 47405-7105, USA
		\and
		INAF -- Osservatorio Astronomico di Padova, Vicolo dell'Osservatorio 5, 35122, Padova, Italy
		\and
		Space Research Institute, Austrian Academy of Sciences, Schmiedlstrasse 6, 8042 Graz, Austria
		\and
		Department of Physics, University of Warwick, Gibbet Hill Road, Coventry, CV4 7AL, UK
		\and
		Centre for Exoplanets and Habitability, University of Warwick, Gibbet Hill Road, Coventry, CV4 7AL, UK
		\and 
		Leiden Observatory, Leiden University, Postbus 9513, 2300 RA Leiden,The Netherlands
		\and
		Istituto di Astrofisica e Planetologia Spaziali INAF-IAPS, Via Fosso del Cavaliere 100, I-00133, Rome, Italy
		\and
		INAF -- Osservatorio Astronomico di Brera, Via E. Bianchi 46, 23807, Merate (LC), Italy
		\and
		INAF -- Osservatorio Astrofisico di Catania, Via S. Sofia 78, 95123, Catania, Italy
		\and
		Department of Physics, University of Rome ``Tor Vergata'', Via della Ricerca Scientifica 1, 00133, Roma, Italy
		\and
		Max Planck Institute for Astronomy, Königstuhl 17, 69117, Heidelberg, Germany
		\and
		INAF -- Osservatorio Astronomico di Palermo, P.zza Parlamento 1, I-90134 Palermo, Italy
		\and
		Dipartimento di Fisica e Astronomia Galileo Galilei, Universit\`a di Padova, Vicolo dell'Osservatorio 3, 35122, Padova, Italy
		\and
		INAF -- Osservatorio Astrofisico di Arcetri, Largo E. Fermi 5, 50125, Firenze, Italy
		\and
		INAF -- Osservatorio Astronomico di Trieste, via Tiepolo 11, 34143, Trieste, Italy
		\and
		Fundaci\'on G. Galilei - INAF (Telescopio Nazionale Galileo), Rambla J. A. Fern\`andez P\`erez 7, 38712, Bre\~na Baja (La Palma), Spain
		\and
		INAF -- Osservatorio Astronomico di Capodimonte, Salita Moiariello 16, 80131, Naples, Italy
		\and
		INAF -- Osservatorio di Cagliari, via della Scienza 5, 09047, Selargius, CA, Italy
		\and
		Dipartimento di Fisica, La Sapienza Università di Roma, Piazzale Aldo Moro 2, 00185 Roma, Italy
	}
	
	\date{Received date ; Accepted date }

	\abstract
 
	{The field of exo-atmospheric characterisation is progressing at an extraordinary pace. Atmospheric observations are now available for tens of exoplanets, mainly hot and warm inflated gas giants, and new molecular species continue to be detected revealing a richer atmospheric composition than previously expected. Thanks to its warm equilibrium temperature (963$\pm$18~K) and low-density (0.219$\pm$0.031~g cm$^{-3}$), the close-in gas giant WASP-69b represents a golden target for atmospheric characterisation.}
	{ With the aim of searching for molecules in the atmosphere of WASP-69b and investigating its properties, we performed high-resolution transmission spectroscopy with the GIANO-B near-infrared spectrograph at the Telescopio Nazionale Galileo.}
	{We observed three transit events of WASP-69b. During a transit, the planetary lines are Doppler-shifted due to the large change in the planet's radial velocity, allowing us to separate the planetary signal from the  quasi-stationary telluric and stellar spectrum. }
	{Considering the three nights together, we report the detection of CH$_4$, NH$_3$, CO, C$_2$H$_2$, and H$_2$O, at more than $3.3\sigma$ level. We did not identify the presence of HCN and CO$_2$ with confidence level higher than 3$\sigma$. This is the first time that five molecules are simultaneously detected in the atmosphere of a warm giant planet. These results suggest that the atmosphere of WASP-69b is possibly carbon-rich and characterised by the presence of disequilibrium chemistry.}
	{}
	
	\maketitle
	\section{Introduction}\label{sec:intro}
	Ground-based, high-resolution (resolving power $R\geq$20\,000) spectroscopy (HRS) in the near-infrared (nIR) represents an effective approach to investigate exoplanet atmospheres. Indeed, at high spectral resolution, thousands of resolved molecular lines can be disentangled, as they are Doppler shifted by tens of km s$^{-1}$ due to the planet motion along its orbit from the telluric (Earth’s atmosphere) and stellar lines that are (quasi-)stationary in wavelength.

	Hot ($T_{\rm eq}>1000$~K) and warm ($T_{\rm{eq}}\leq1000$~K) inflated giant planets are the most suitable targets for atmospheric characterisation \citep{Seager2010}. Recently, \citet{Giacobbe2021} reported the simultaneous detection of six molecules in the atmosphere of one of the best-studied hot-Jupiters (HJs), HD\,209458b. As only two molecules (CO and H$_2$O) had been detected at the same time in exoplanetary atmospheres in the past \citep[e.g.,][]{Tsiaras2018}, the work presented in \citet{Giacobbe2021} revealed a previously unknown chemical complexity.
	Thus this discovery raises the question of whether the complexity of HD\,209458b's atmosphere is unique or other exo-atmospheres show such a rich molecular composition. %arises spontaneously. 
	Such molecular diversity enables one to derive atmospheric abundances and thus elemental ratios, such as Carbon-to-Oxygen (C/O) ratio, can then be obtained, allowing us to derive important clues on the formation and migration histories of hot giant planets \citep[e.g.,][]{Madhusudhan2012apjCOratios, Turrini2021a}. For example \citet{Giacobbe2021} retrieved a C/O ratio $\geq$ 1, meaning that HD\,209458b formed far from its present location and subsequently migrated inwards.\\
	\indent With a warm temperature and low density (see Table~\ref{Table1}), the close-in transiting planet WASP-69b represents a suitable laboratory for atmospheric characterization studies. Furthermore, the low equilibrium temperature T$_\mathrm{eq}\leq$1000~K places the planet in a different regime compared to that of the HJ HD\,209458b (T$_{\rm{eq}}\sim$1500~K): the equilibrium chemistry shifts in favour of CH$_4$ rather than CO and disequilibrium phenomena (i.e. vertical mixing and photochemical processes) are more likely to alter the molecular abundances compared to the atmospheres of hotter planets \citep[e.g.,][]{Moses2014}.
	
	The atmosphere of WASP-69b has been studied with both space-based low-resolution (LR) and ground-based high-resolution (HR) in-transit (transmission spectroscopy) observations. LR observations performed with WFC3 on board the Hubble Space Telescope (HST) unveiled the presence of H$_2$O and possibly aerosols \citep{Tsiaras2018,Fisher2018}. Sodium has also been detected through HR optical observations \citep[e.g.,][]{Casasayas2017}. Moreover, HR and LR observations covering the nIR \ion{He}{I} (1083.3~nm) revealed that the planet has an extended and escaping atmosphere, where the escape is driven by the intense X-ray and extreme ultraviolet radiation received from the active K5V host star \citep[e.g.,][]{Nortmann2018, Vissapragada2020}.

\begin{figure}
		\centering
		\includegraphics[width=\linewidth]{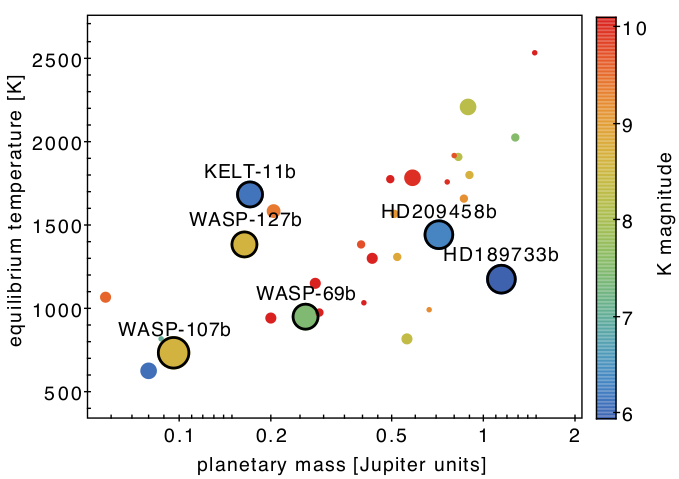}
		\caption{
			Mass-equilibrium temperature relation for the planets within the GAPS atmospheres survey. The symbol size is proportional
			to the expected atmospheric signal (for one scale height) in K band.The six more suitable targets to perform atmospheric characterisation studies are labelled.}
		\label{Valerio_plot}
	\end{figure}

	In this work, we present transmission spectroscopy of WASP-69b using HR observations in the nIR, searching for the absorption of multiple molecular species. We observed the WASP-69 system within the long-term observing program at the Telescopio Nazionale Galileo (TNG) telescope ``GAPS2: the origin of planetary systems" -awarded to the Italian GAPS (Global Architecture of Planetary System) Collaboration \citep{Poretti2016}. Part of this program is devoted to the characterisation of exoplanetary atmospheres at HR. 
	Briefly, we are carrying out transmission and emission spectroscopy of 26 hot Jupiters and 4 hot Neptunes and/or sub-Neptunes in a wide range of T$_{\rm{eq}}$ (625-2500~K), and hosted by stars of different spectral types (A to M with $K < 10.5$\,mag) to ($i$) identify atomic species to estimate the abundance of refractory elements \citep{Pino2020} and/or to probe atmospheric expansion/escape, for instance from the H$\alpha$ \citep{Borsa2021} and/or metastable Helium triplet \citep{Guilluy2019}; ($ii$) detect molecular compounds  and estimate the atmospheric C/O ratio and metallicity, which in turn provide constraints on the planet formation and evolution scenarios \citep{Giacobbe2021}; ($iii$) derive temperature-pressure profiles \citep{Pino2020, Borsasubmitted}; ($iv$) detect the atmospheric Rossiter-McLaughlin effect \citep{Borsa2019,Rainer2021}.\\
	Fig.~\ref{Valerio_plot} places WASP-69b into the broader context of the GAPS atmospheric survey. The symbol size is proportional to the expected atmospheric signal in the K band:
	\begin{equation}
		S/N =\frac{2HR_P}{R_\star^2}*\sqrt{F},
	\end{equation}
	where we account for the stellar magnitude in K band ($F=10^{-\frac{magK}{2.5}}$), the atmospheric scale height ($H$), the planetary ($R_P$) and the stellar radius ($R_\star$). The most suitable targets to perform atmospheric characterisation studies are labelled. As this plot shows, WASP-69b represents one of the best suitable candidates to perform atmospheric studies (2R$_\mathrm{P}$H/R$_\star^2 \sim$ 283~ppm\footnote{For comparison, HD~209458~b investigated in \citet{Giacobbe2021} has a 2R$_\mathrm{P}$H/R$_\star^2 \sim$ 175 ppm.}, \citealt{Brown2001} ) in the temperature range ($\leq$1000~K) we aimed to probe within this study.
	
	\begin{table*}[h]
		\centering
		%\footnotesize
		\caption{Physical and orbital parameters of the WASP-69 system adopted in this work.}

		\begin{threeparttable}
			\label{Table1}
			\begin{tabular}{l c c}
				
				\hline \hline
				%\small
				\textbf{Orbital parameters}  & & \\
				\hline
				Orbital Period - P[days]	&		3.8681382(17)            &   		1,2\\
				Mid Transit time - T$_\mathrm{c}$[BJD$_\mathrm{TDB}$]   &      		2455748.83422(18)     &       	1,2\\
				Semi-major axis -  a[au]&    	            	0.04527$^{+0.00053}_{ -0.00054}$ &		1\\
				Orbital inclination - i[deg]       &              		86.71(20)       &                          		1,2\\
				Systemic velocity - Vsys[km/s]     &           	-9.37(21)                 &                  		3\\
				Transit Duration - T$_\mathrm{14}$[h]    &      		2.2296(288)                   &         		2\\
				Radial-velocity semi-amplitude - K$_\mathrm{P}$[km/s]      &            	 127.11$^{+1.49}_{ -1.52}$      &          		 This paper\tnote{*}\\
				Orbital eccentricity - e & < 0.11 & 1\\
				Transit ingress/egress - $\phi_1$ & 0.01201(16) & This paper\tnote{°}\\
				\hline
				\textbf{Planetary Parameters}  &  & \\
				\hline
				Mass -   M$_\mathrm{P}$[M$_\mathrm{Jup}$] & 0.260(17) & 2 \\
				Radius -  R$_\mathrm{P}$[R$_\mathrm{Jup}$] & 1.057(47) & 2 \\
				Equilibrium Temperature -  T$_\mathrm{eq}$[K] & 963(18) & 2 \\
				Density -  $\rho_\mathrm{P}$[kg m$^{-3}$] & 219000(31000)  & 2 \\
				\hline
				\textbf{Stellar Parameters}  &  &\\
				\hline   
				Mass - M$_\mathrm{\star}$[M$_\sun$] & 0.826(29) & 2 \\
				Radius - R$_\mathrm{\star}$[R$_\sun$] & 0.813(28) & 2 \\
				Effective Temperature - T$_\mathrm{\star}$[K] & 4715(50) & 2 \\
				Spectral Type  & KV5 & 2 \\
				Index color - B-V & 1.06 & 4 \\
				Stellar Rotational velocity - vsini[km/s] & 2.20(40) & 1,2 \\
				\hline
			\end{tabular}%}
			\begin{tablenotes}[flushleft]
			\item 	$^*$Derived from $a$, $P$, and $i$ as $\frac{2 \pi a}{P}\sin{i}$,\\
				$^+$Derived from HARPS-N,\\
					$^{o}$Orbital phase corresponding to transit ingress/egress derived as $\frac{T_{14}}{2\,P}$
				\item \textbf{References: } $^{(1)}$\citet{Bonomo2017}, $^{(2)}$\citet{Anderson2014}, 
				$^{(3)}$\textit{Gaia} DR2 \citep{Gaiadr2}, $^{(4)}$From the Tycho-2 catalogue \citep{Hog2000},
						
			\end{tablenotes}
		\end{threeparttable}
	\end{table*}
	This letter is organized as follows: in Sect.~\ref{data_sample} we present the data we collected, in Sect.~\ref{data_analysis} we describe
	the analysis that we performed to extract the planetary transmission spectrum and, therefore, the molecules responsible for absorption, from the raw GIANO-B data. Finally, we discuss our results and draw the conclusions in Sect.~\ref{results}.
	\begin{figure}
		\centering
		\includegraphics[width=\linewidth]{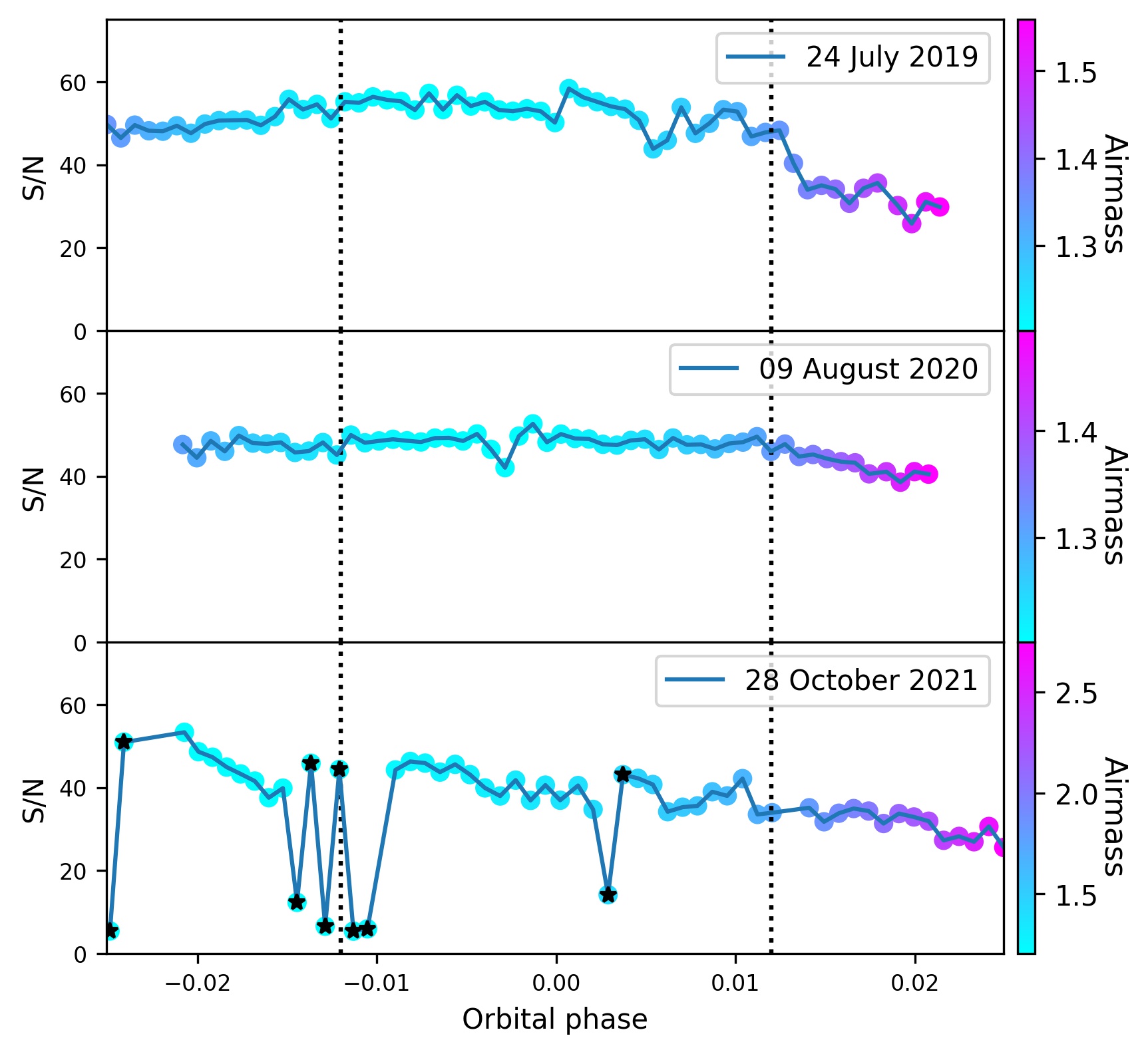}
		\caption{Average S/N across the entire GIANO-B spectral range as a function of the orbital phase, with the marker color indicating the airmass at which the observations have been gathered. Ingress and egress of WASP-69b ($\phi_1$ in Table~\ref{Table1}) are marked with vertical dashed lines. Black stars on the last night (October 28, 2021) indicate the discarded AB couples for low S/N. }
		\label{snr_ph_am}
	\end{figure} 
	
	\section{Data sample} \label{data_sample}
	We observed the WASP-69 system with the TNG telescope in GIARPS configuration by simultaneously measuring HR spectra in the optical (0.39-0.69~$\mu$m) and nIR (0.95-2.45~$\mu$m) with the HARPS-N ($R \approx 115\,000$) and GIANO-B (R$\sim$50,000) spectrographs. Our data encompass a total of three primary transits of WASP-69\,b, which were obtained at UT 24 July 2019, UT 09 August 2020 and UT 28 October 2021. 
	The GIANO-B observations were taken according to an ABAB nodding pattern \citep{GIARPS_claudi} for an optimal subtraction of the thermal background noise and telluric emission lines. The observations consist of spectra taken before, during, and after the planetary transit; Fig.~\ref{snr_ph_am} shows the signal-to-noise ratio (S/N) averaged over the entire GIANO-B spectral coverage as a function of the planet's orbital phase and of the airmass for each of the three nights. The observations during the last night (October 28, 2021) were affected by several thin clouds (cirri), so we decided to discard the AB couples of observations which exhibit a very low S/N compared to the others (lowest S/N in the couple less then 15, see Fig.~\ref{snr_ph_am}).
	A log of the observations is reported in Table~\ref{log}.
	\begin{figure*}
		\centering
	
		\includegraphics[width=\linewidth]{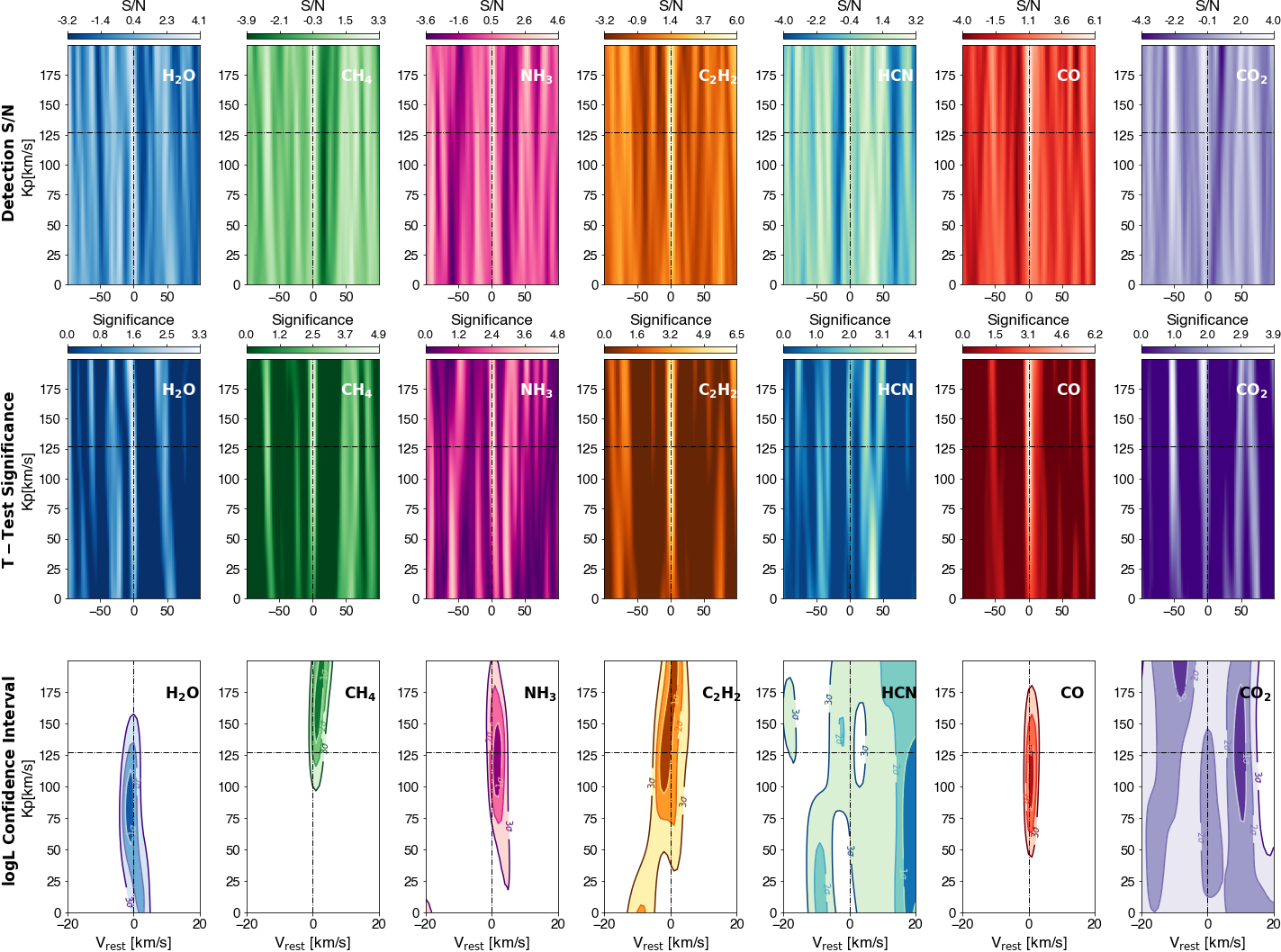}
		\caption{CC both in the S/N (top panel) and in the significance (middle panel) framework, and LH confidence intervals (bottom panel) maps for each investigated molecules as a function of the planet’s maximum radial velocity ($K_{\mathrm{P}}$) and the
			planet’s rest-frame velocity ($V_{\mathrm{rest}}$). The  dotted lines denote the expected radial-velocity semi-amplitude of WASP-69b ($K_{\mathrm{P}}$=127.11$^{+1.49}_{ -1.52}$ km~s$^{-1}$) in the planet rest-frame ($V_{\mathrm{rest}}$=0 km~s$^{-1}$). The detection significance increased when transits are co-added.}
		\label{sigma}
	\end{figure*}
	\section{Data Analysis} \label{data_analysis}
	Here we report the steps we performed in order to recover the planetary signal from the GIANO-B raw images.
	
	\textbf{($i$) Basic data reduction}.
	We processed the raw GIANO-B spectra with the GOFIO pipeline \citep{Rainer2018} to perform flat-fielding, bad pixel correction, background subtraction, optimal extraction of the 1-d spectra, and a preliminary wavelength calibration using an Uranium-Neon lamp. We corrected small temporal variations in the wavelength solution due to non-perfect stability of the spectrograph \citep{Giacobbe2021}, and refined the initial wavelength solution by employing a HR transmission spectrum of the  Earth’s atmosphere generated via the ESO Sky Model Calculator\footnote{\url{https://www.eso.org/observing/etc/bin/gen/form?INS.MODE=swspectr+INS.NAME=SKYCALC}}. As some spectral regions have too many or too few telluric lines, this calibration refinement was not possible for all the orders \citep[e.g.,][]{Brogi_2018_Giano, Guilluy2019, Fossati2022}. We report in Table~\ref{bad_orders} the discarded orders for each individual night of observation. 
	
	\textbf{($ii$) Telluric and stellar Lines Removal}.
	We then removed the telluric contamination and the stellar spectrum by employing the Principal Component Analysis (PCA) and linear regression techniques \citep{Giacobbe2021}. We selected the optimal number of components $N_\mathrm{opt}$ following the procedure described in \citet{Giacobbe2021}. More precisely, $N_\mathrm{opt}$ varies with the considered order depending on the Root Mean Square (rms) variation. We selected the number for which the first derivative between the last and penultimate component decreases by less than $\sigma_\mathrm{white}N_\mathrm{opt}^{-1/2}$, where $\sigma_\mathrm{white}$ is the standard deviation of the full matrix assuming pure white noise.  Since in two nights out of three (i.e., UT 24 July 2019, and UT 09 August 2020) the relative velocity shift between telluric and planetary lines was close to zero, we masked the telluric components to avoid contamination in our final spectra. Following \citet{Brogi_2018_Giano}, we thus masked the data assigning
	zero flux to those spectral channels corresponding to telluric
	lines deeper than 20$\%$. Finally, we visually inspected the outcome for spurious telluric residuals.\\

	\textbf{($iii$) Search for molecules through cross-correlation}. For each considered molecule, we combined the contribution of thousands of lines in the spectral coverage of GIANO-B by performing the cross-correlation  (CC) of the residual spectra with isothermal transmission models generated using the GENESIS code \citep{Gandhi2017} adapted for transmission spectroscopy \citep{Pinhas2018} (see Table~\ref{table_res} for the adopted line list). Prior to CC, the models were convolved to the GIANO-B instrument profile (a Gaussian profile with FWHM~5.4\,km\,s$^{-1}$). Models were calculated between 100 and 10$^{-8}$~bar in pressure, at a T$_\mathrm{eq}=963(18)$K, between 0.9 and 2.6$\mu$m, with a constant spacing of 0.01~cm$^{-1}$. We included collision-induced absorption from H$_2$-H$_2$ and H$_2$-He interactions \citep{RICHARD2012}.
	For each night we performed the CC for an optimal selection of GIANO-B orders \citep{Giacobbe2021} and every phase over a lag vector corresponding to planet RVs in the range $-252 \leq {\rm RV} \leq 252$\,km\,s$^{-1}$, in steps of 3~km~s$^{-1}$. This range was chosen in order to cover all possible radial velocities of the planet.
	We then co-added the CC functions over the selected orders, nights and the orbital phase after shifting them in the planet rest-frame by assuming a circular orbit \citep[e.g.,][]{Brogi2012, Brogi_HD179949b_2014,Brogi_2018_Giano}.
		In doing so, we considered a range of planet radial-velocity semi-amplitudes $ 0 \leq K_{\rm p} \leq 200$\,km\,s$^{-1}$, in steps of 1.5~km~s$^{-1}$. 
	We then quantified the confidence  level  of  our detection by computing the S/N detection map \citep[e.g.,][]{Brogi2012, Brogi_HD179949b_2014,Brogi_2018_Giano}. For each investigated molecule, we divided the total CC matrix by its standard deviation (calculated by excluding the
	CC peak). We also used the Welch t-test to compare the  ‘in-trail’  values (CC values that carry signal) of  the  2-d CC matrix  aligned in the planet rest-frame to those ‘out-of-trail’ (CC values away from the WASP-69b's radial velocity). We then converted the corresponding t-value into the statistical significance at which these two distributions are not drawn from the
	same parent distribution. We applied this method for the same range of rest-frame velocity $V{_\mathrm{rest}}$ and the planetary semi-amplitude $K{_\mathrm{P}}$ adopted before. The final S/N and significance maps for the three nights combined together are shown in the upper and in the middle panels of Fig.~\ref{sigma}, respectively. These results show the advantage of a multi-night approach, indeed even if the data quality does not allow to always detect the signal in each individual night (see Fig.~\ref{single_night}), the detection significance increased when transits are co-added. \\
	\indent The detections in CC have been obtained with isothermal models of one molecule at a time with a Volume Mixing Ratio (VMR) large enough to reveal a given molecule, if present in the atmosphere; it has been indeed demonstrated \citep[e.g.,][]{Gandhi2020b} that the CC is only weakly dependent on the absolute VMR. As a good compromise we found a VMR$\sim$10$^{-4}$.

	\textbf{($iv$) Confidence Intervals via Likelihood}.
	The CC framework as applied in the previous analysis implicitly assumes that the signal amplitude and the S/N are uniform across the considered spectral range. However, this assumption is only valid at a first-order level. To investigate this aspect more thoroughly, we converted the CC values into Likelihood (LH) Mapping values using the approach proposed by \citet{Brogi2019} and used by \citet{Giacobbe2021}. In this formalism, the LH is indeed set up by taking into account the model's line depth and the S/N order by order and spectrum by spectrum. Retrieving a peak in the LH at the correct position of the ($V_{\mathrm{rest}}$, $K_{\mathrm{P}}$) space is hence an additional confirmation of our detection.
	We thus computed a log-LH function for each selected order, each spectrum, and each radial-velocity shift of the model across the $K_{\mathrm{P}}$ vs $V_{\mathrm{rest}}$ space. We used a lag vector corresponding to planet radial velocities (RVs) in the range $-20 \leq {\rm RV} \leq 20$\,km\,s$^{-1}$, in steps of 1~km~s$^{-1}$. To take into account any possible modification of the planetary signal due to the PCA process, we performed on each theoretical model the exact telluric removal process applied to the observations.  
	The final log-likelihood for each ($V_{\mathrm{rest}}$, $K_{\mathrm{P}}$) couple is obtained by adding all the log-LHs over each order, each night, and each observed spectrum. %In this way, for each molecule we obtained a likelihood map in the ($V_{\mathrm{rest}}$, $K_{\mathrm{P}}$) space.
	
	Subsequently, as in \citet{Giacobbe2021}, we introduced the line-intensity scaling factor $S$ ($-1.1$<$\log{S}$<$1.1$, in steps of 0.1), to understand how much the used models have to be scaled to match the observed data. 
	For a model perfectly matching the data, $S$ is exactly 1 ($\log{S}$=0), i.e. the model and
	the data have the same average line amplitude compared to the local continuum. 
	Scaling factors smaller than 1 indicate that spectral lines are too strong compared to
	the observations and, for example, the presence of a grey-cloud model can be invoked to account for that, as in \citet{Giacobbe2021}. Furthermore, as in the single species models the continuum can be approximate, the introduction of the $\log{S}$ allows us to account for the muting effects of other species. On the contrary in the CC framework, as the line depth does not enter in the calculation, the $\log{S}$ is it not necessary. 
	We then converted the LH values into confidence intervals by using the LH-ratio test between the maximum log-LH and each LH value in the ($V_{\mathrm{rest}}$, $K_{\mathrm{P}}$, $\log{S}$) space. The LH confidence intervals maps corresponding to the
	$\log{S}$ maximising the detection, are shown in the bottom panel of Fig.~\ref{sigma}.

	We then evaluated the significance of our LH detection by performing a LH-ratio test between the maximum log-LH and the median of the 2-d log-LH matrix values for the best-fit scaling factor which exhibited a statistical confidence greater than 1. For each considered molecule, Table~\ref{table_res} reports the results of our analysis in both the CC and the LH frameworks. We consider a molecule as detected if altogether the S/N, the detection significance and the LH significance are higher than 3$\sigma$. 

	\begin{table*}[ht]
		\centering
		\footnotesize
		\caption{CC and LH results.}
		\resizebox{0.99\textwidth}{!}{
			\begin{threeparttable}
				
				\label{table_res}
				% %
				\begin{tabular}{c | l | c | c c c | c c c | c c c c }
					\hline
					\hline
					& & & \multicolumn{6}{|c|}{\textbf{CC framework}} & \multicolumn{4}{c}{\textbf{LH framework}} \\ \cline{4-13}
					
					& & & \multicolumn{3}{c}{\textbf{S/N map}} & \multicolumn{3}{|c|}{\textbf{Significance map}} & \multicolumn{4}{c}{\textbf{Confidence Intervals map}} \\
					\textbf{Mol}& \textbf{Line list} & \textbf{Status} & \textbf{$\mathbf{V{_{\mathrm{\mathbf{rest}}}}_0}$ [km/s]} & \textbf{Kp$_0$ [km/s] } & \textbf{S/N} & \textbf{$\mathbf{V{_{\mathrm{\mathbf{rest}}}}_0}$ [km/s]} & \textbf{Kp$_0$ [km/s] } & $\sigma$ & \textbf{$\mathbf{V{_{\mathrm{\mathbf{rest}}}}_0}$ [km/s]} & \textbf{Kp$_0$ [km/s] } & \textbf{$\log{S}$} & \textbf{$\sigma$}   \\
					\hline
					H$_2$O & POKAZATEL/ExoMol\tablefoottext{a}& \cmark & 0.0$^{+ 3.0}_{-3.0}$ & 114.0$^{+ 58.5}_{- 57.0}$ &  4.1 & 0.0$^{+ 3.0}_{-3.0}$ &111.0$^{+ 67.5}_{- 82.5}$ &  3.3 & -1.0$^{+ 1.0}_{-2.0}$& 84.0$^{+ 30.0}_{- 31.5}$ & -0.6 & 5.0 \\
					CH$_4$  & HITEMP\tablefoottext{b}& \cmark &   0.0$^{+ 3.0}_{-3.0}$ &130.5$^{+ 69.0}_{- 67.5}$ &  3.3 & 0.0$^{+ 3.0}_{-3.0}$ & 132.0$^{+ 67.5}_{- 60.0}$ &  4.9 & 2.0$^{+ 1.0}_{-1.0}$ &171.0$^{+ 29.0}_{- 30.0}$ & -0.1 & 6.3 \\
					NH$_3$ & CoYuTe/ExoMol\tablefoottext{c}& \cmark &   0.0$^{+ 3.0}_{-3.0}$ &148.5$^{+ 51.0}_{- 57.0}$ &  4.6 & 0.0$^{+ 3.0}_{-3.0}$ &154.5$^{+ 45.0}_{- 58.5}$ &  4.8 & 2.0$^{+ 1.0}_{-2.0}$ & 121.5$^{+ 30.0}_{- 31.5}$ & -0.3 & 5.4 \\
					C$_2$H$_2$ &  ACeTY/ExoMol\tablefoottext{d} & \cmark  &   0.0$^{+ 3.0}_{-3.0}$ &130.5$^{+ 52.5}_{- 52.5}$ &  6.0 & 0.0$^{+ 3.0}_{-3.0}$ &129.0$^{+ 60.0}_{- 52.5}$ &  6.5 & -2.0$^{+2.0}_{- 1.0}$ &123.0$^{+ 33.0}_{- 30.0}$ & -0.2 & 4.5 \\
					HCN & Harris/ExoMol\tablefoottext{e}& \xmark & & & & & & & & & &  \\
					CO   & HITEMP\tablefoottext{f}& \cmark & 0.0$^{+ 3.0}_{-3.0}$ &148.5$^{+ 51.0}_{- 69.0}$ &  6.1 & 0.0$^{+ 3.0}_{-3.0}$ &165.0$^{+ 34.5}_{- 81.0}$ &  6.2 & 1.0$^{+ 1.0}_{-1.0}$&108.0$^{+ 28.5}_{- 28.5}$ &  0.1 & 7.0  \\
					CO$_2$  & Ames\tablefoottext{g} & \xmark &  &   & &  &   &   &  &  &  &  \\
					\hline
					
				\end{tabular}%}
				\begin{tablenotes}[flushleft]
					\item \textbf{Notes: }From left to right: the investigated molecule, the adopted line list, the CC result with theoretical models both in the S/N and the significance framework, the LH findings, and the status of the detection (with \cmark and \xmark, indicating a detection and a non-detection, respectively). For each framework, the planet orbital semi-amplitude (\textit{Kp$_0$}), the velocity in the planet rest-frame (\textit{$V{_{\mathrm{rest}}}_0$}) of the CC/LH peak are reported. The S/N, the significance ($\sigma$), and the scaling factor ($\log{S}$) maximising the detection are also present. We used a grid of $\log{S}$ values, thus they are shown without an error bar.
					\item \textbf{References: }\tablefoottext{a}{\citet{polyansky_h2o}},\tablefoottext{b}{\citet{CH4_hitemp}}, \tablefoottext{c}{\citet{nh3}}, \tablefoottext{d}{\citet{acety}}, \tablefoottext{e}{\citet{HCN}}, \tablefoottext{f}{\citet{COandCO2,li_co_2015}}, \tablefoottext{g}{\citet{Ames}}
				\end{tablenotes}
		\end{threeparttable}}
	\end{table*}
	
	\begin{figure*}[h]
		\centering
		\includegraphics[width=17cm, clip, trim=5 5 5 5]{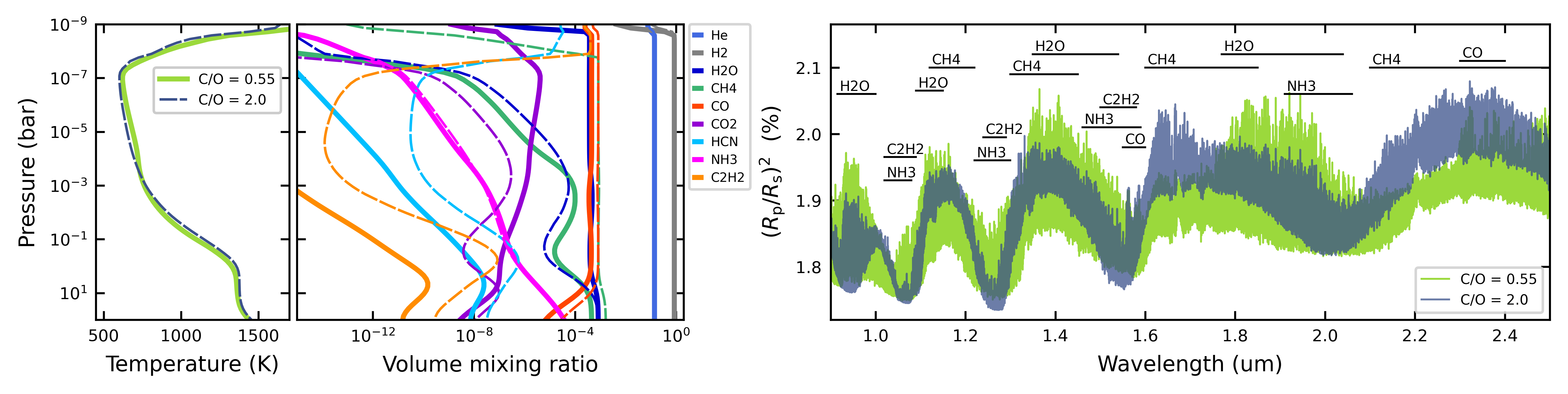}
		\caption{Sample atmospheric models for WASP-69b in radiative and
			thermochemical equilibrium.  The left panel shows the
			equilibrium temperature profile for two models with solar elemental
			metallicity and a solar C/O ratio (solid green curve) or a
			super-solar ratio (dashed blue curve).  The middle panel shows
			the equilibrium composition for the respective solar C/O model
			(solid curves) and super-solar C/O model (dashed curves), colour
			coded for each relevant species (see legend). The right panel shows
			the post-produced HR transmission spectra over the GIANO-B wavelength range, for the
			solar C/O model (green) and super-solar C/O model (blue).  The
			horizontal black lines denote the wavelength ranges where each of
			the labelled species significantly impact the spectra.}
		\label{fig:atm_physics}
	\end{figure*}
	%\end{comment}
	\section{Results and Discussion} \label{results}
	We characterise the atmosphere of WASP-69b showing a multi-species molecular detection in the planet transmission spectrum. Our results indicate the presence of CH$_4$, NH$_3$, CO, C$_2$H$_2$, and H$_2$O, at more than $3.3\sigma$ level (Table~\ref{table_res}). We did not detect the presence of HCN and CO$_2$ with confidence level higher than $3\sigma$ (see Table~\ref{table_res}). The retrieved $K_{\rm p}$ values for all molecules are compatible within $1\sigma$ with the nominal $K_{\rm p}$ value in the S/N and t-test maps (see Table~\ref{table_res} and Fig.~\ref{sigma}, top and middle panels). We find the same occurring for NH$_3$, CO and C$_2$H$_2$ when switching to the LH framework, while the peaks of the CH$_4$ and H$_2$O signals are slightly shifted, though consistent with the expected $K_{\rm p}$ at the $2\sigma$ level (see Fig.~\ref{sigma}, bottom panel). Since the LH function is more sensitive to the variance of the spectra, these small shifts may be due to the contribution of the other molecules, which are not modelled simultaneously when considering one molecule at a time.  	
	We also checked whether our CO detection might be spurious, as stellar
	CO lines can still be an important contaminant of the planetary transmission
	spectrum \citep{Brogi2016}. To resolve the possible ambiguity, we
	evaluated the expected $K_\mathrm{P}$ at which a spurious signal of stellar origin
	should have been maximised as $K{_\mathrm{P}}_\mathrm{star}\sim\frac{vsini}{\sin(2\pi \phi_1)}\sim$29.2$\pm$5.3~km/s (by using the values in Table~\ref{Table1}). % is not
	While the above equation provides an estimate of the expected position of stellar residuals in the (V$_\mathrm{rest}$,$K_\mathrm{P}$) map, the actual position will depend on the interplay between the stellar and the planetary signal, and might also be affected by any partial removal of stellar residuals via PCA. Thus, one should not expect stellar residuals to arise exactly at the $K{_\mathrm{P}}_\mathrm{star}$  computed above. In fact, there are cases in the literature where stellar CO causes the planet signal to be detected at higher $K{_\mathrm{P}}$ than the orbit of the planet allows \citep[e.g.,][Fig.7]{Chiavassa2019}, which seems at first counter-intuitive.\\
	\indent With the LH framework applied, we could also infer to which extent the model is a plausible representation of the data. 
	However, we measured low $\log{S}$ values for several molecules, as single-species models have higher variance (deeper lines) than an actual spectrum, because they do not contain the additional opacity from other species.
	We remark that the assumption of fixed VMR composition has an impact on the line strengths $S$, and can thus lead to ambiguities in the scaling factors interpretation. Consequently, the comparison between the $S$ values for the different molecules is not straightforward, meaning that the relative abundances between the different species cannot be easily estimated. The scaling factors obtained here (Table~\ref{table_res}) must thus be considered representative and taken with caution.

	The large number of individual species detections (and non-detections) on
	WASP-69b provides important constraints on the physical processes that
	govern the planet’s atmosphere. Using the \textsc{Pyrat Bay} \citep{CubillosBlecic2021mnrasPyratBay} modeling framework (see Appendix Sect.~\ref{pyratbay}), we
	explored physically plausible atmospheric scenarios in radiative and
	thermochemical equilibrium that could be consistent with the observed
	molecular detections.
	Given the striking difference in composition that is expected for
	scenarios with elemental ratios of C/O<1 or C/O>1 \citep[see,
	e.g.,][]{Tsuji1973aaStellarAtmospheres, Madhusudhan2012apjCOratios,
		MosesEtal2013apjChemistryOfCOratios}, we adopted the C/O ratio and
	the atmospheric metallicity as the main proxies to explore different
	plausible compositions for WASP-69b. We found that very high-metallicity regimes ($\gtrsim 10\times$ solar)
	appear disfavoured by the observations because, 
	for C/O<1 scenarios, the non-detected {\carbdiox} molecule would present detectable features since its abundance increases faster with metallicity than for the other species. For C/O>1 scenarios at high metallicities, the observed {\water} features would not be detected since they would be obscured by the {\methane} absorption, due to the much lower abundance of {\water} relative to {\methane}. When comparing solar metallicity models with C/O ratios lower and
	greater than one (Fig.~\ref{fig:atm_physics}), the observations seem
	to be more consistent with the latter. First, while transmission
	models for both regimes show clear spectral features of {\water},
	{\methane}, and CO, the stronger detection of {\methane} than {\water}
	may be favouring the C/O>1 scenario (in which {\methane} is
	predominant).
	Then, the {\ammonia} molecule presents wide-range but weak features
	across the spectrum in both C/O regimes, thus, it does not
	particularly favour any of the scenarios.  However, the {\acetylene} is hard to explain in equilibrium conditions, since its estimated abundance does not yield detectable features in any of the above scenarios. 
	A way to resolve this conundrum is by including disequilibrium
	chemistry \citep[see e.g.,][and references therein]{Moses2014}, which
	would significantly enhance the abundances of {\ammonia} and
	{\acetylene} by dragging material from the deep interior to the upper
	atmospheric layers (vertical quenching), or by photochemical production,
	or both.
	Certainly, the much greater abundance of {\acetylene} for the C/O>1
	case makes the detection of this molecule more plausible for this
	scenario than for C/O<1 (a difference of $\sim$5 orders of magnitude
	for the cases shown in Fig.~\ref{fig:atm_physics}), as long as
	disequilibrium processes help to bridge the gap between the WASP-69b
	detections and the equilibrium theoretical predictions.

	To give a more quantitative interpretation of the physical and chemical conditions of WASP-69b’s atmosphere, we compute and test three different scenarios that could be a possible representation of WASP-69b's atmosphere:
	\begin{itemize}
		\item (i) solar metallicity and solar C/O (C/O=0.55)
		\item (ii) solar metallicity and carbon rich (C/O=2.0)
		\item (iii) solar metallicity, carbon rich (C/O=2.0), and quenched C$_2$H$_2$ and NH$_3$ to VMRs of $10^{-6}-10^{-7}$.
	\end{itemize}
	As opposed to the single-molecule analyses from Section~\ref{data_analysis}, here we
	generate more realistic "full spectrum" models combining opacities
	from all species. To quantify which scenario describes better the observations, we apply
	the Wilks theorem \citep{wilks} by computing the LH-ratio test relative to the LH-maxima (following the procedure of \citealt{Giacobbe2021}).
	Our analysis (see Table ~\ref{lmax}) indicates that scenario (\textit{iii}) is marginally ($\sim 1 \sigma$) preferred compared to scenario (\textit{ii}), and significant favoured ($\sim 3.5 \sigma$) over scenario (\textit{i}) . The atmosphere of WASP-69 seems thus to be better described by a scenario where quenching of NH$_3$ and CH$_4$ is taking place. However, we need to stress two aspects: (1) this result does not imply that (\textit{iii}) is the best-fitting scenario, but it is the preferred one within the tested models; and (2) we assumed a fixed solar metallicity, which might not be the most representative of the planet's atmosphere. Dealing with (1) and (2) implies a retrieval analysis that goes beyond the scope of this paper.

	The emerging picture provides some preliminary indications on WASP-69b's formation history once put in the context of the host star. WASP-69's metallicity is super-solar \citep{Anderson2014}, with recent works suggesting metallicity values 2-3 higher than that of the Sun \citep{Magrini2022}. The atmospheric scenario favoured by our modelling efforts points to a solar metallicity for WASP-69b's atmosphere, meaning that the atmospheric metallicity is sub-stellar. The combination of sub-stellar planetary metallicity and planetary C/O>1 is the signpost of giant planets whose content of heavy elements is dominated by the accretion of gas with limited contribution of solids \citep{Turrini2021b,Turrini2021a}. Furthermore, this combination suggests that the chemical structure of WASP-69b's native circumstellar disc was inherited from the parent molecular cloud \citep{Pacetti2022}.\\
	We also note that cloud formation processes offer an alternative interpretation of these results. In an atmosphere with prominent cloud formation, condensation locks and transports heavy elements like carbon or oxygen across an atmosphere, which can significantly modify the local gas-phase carbon-to-oxygen ratios.  For example, for a solar-like composition, oxygen depletion due to condensation can be strong enough to raise the carbon-to-oxygen ratios up to ${\rm C/O} \approx 0.7$ \citep{BilgerEtal2013mnrasCloudFormation}.  For carbon-rich atmospheres, on the other hand, carbon depletion can make an atmosphere locally more oxygen rich, approaching values of ${\rm C/O} \approx 1.0$ at the regions with the strongest carbon depletion \citep{HellingEtal2017aaCarbonCloudFormation}.
	
	\indent Whereas the picture described above will need to be validated by future studies, the simultaneous presence of a handful of chemical species constitutes a very important step forward, as so far only a few molecules have been firmly detected simultaneously in the atmosphere of warm giant planets \citep[e.g.,][]{Tsiaras2018, Fisher2018}. Our findings reinforce the notion that a rich atmospheric chemistry is not the sole dominion of hotter planets such as HD\,209458b \citep{Giacobbe2021}. 
	The frontier in the characterisation of exoplanetary atmospheres is expanding further, with even better prospects in the future when stronger constraints on molecular abundances, and hence on metallicity and elemental ratios \citep[e.g.,][]{Brogi2017, Brogi2019, Guilluy2022}, will be placed based on systematic joint analyses of HR ground-based observations gathered with e.g. GIANO-B, CARMENES, Spirou, NIRPS, and CRIRES+, and LR space-borne spectra such as those obtained with HST, JWST, and Ariel.

	\bibliographystyle{aa}
	\bibliography{ref69}	
	
	\begin{acknowledgements}
		We thank the referee, Dr.~Michael Line for his insightful comments which helped to improve the quality of our work. We acknowledge financial contributions from PRIN INAF 2019, and from the agreement ASI-INAF number 2018-16-HH.0 (THE StellaR PAth project). P. C. was funded by the Austrian Science Fund (FWF) Erwin Schroedinger
		Fellowship J4595-N. D.T., E.S. and E.P. acknowledge the support of the Italian National Institute of Astrophysics (INAF) through the INAF Main Stream project “Ariel and the astrochemical link between circumstellar discs and planets” (CUP: C54I19000700005), and of the Italian Space Agency through the ASI-INAF contract no. 2021-5-HH.0. E.P. also acknowledges support from the European Research Council via the Horizon 2020 Framework Programme ERC
		Synergy “ECOGAL” Project GA-855130. M.B. ackowledges support from from the UK Science and Technology Facilities Council (STFC) research grant ST/T000406/1.

	\end{acknowledgements}
	
	\begin{appendix}

		\section{Additional Figures and Tables} \label{add}

		\begin{table}[h]
			\caption{Log of the WASP-69b GIANO-B observations.} 
			\resizebox{0.99\linewidth}{!}{
				\begin{threeparttable}
					
					\label{log}
					\footnotesize
					\begin{tabular}{c  c  c  c c c}
						\hline \hline
						\textbf{Date{$^*$}} & \textbf{N$_{\mathrm{obs}}$} & \textbf{Exp Time} & \textbf{S/N$_{\mathrm{avg}}$} & \textbf{S/N$_{\mathrm{min}}$-S/N$_{\mathrm{max}}$} & \textbf{Airmass} \\
						\hline
						2019-07-24 & 60 & 200~s & 48 & 6 - 90  & 1.2-1.6\\
						2020-08-09 & 54 & 200~s & 47 &  6 - 85 & 1.2-1.5\\
						2021-10-28 & 56 & 200~s & 35 & 3 - 79 & 1.2- 2.8\\
						\hline
					\end{tabular}
					%{ \tablefoottext{$\mathrm{1}$}{Start of night date}.}
					
					\begin{tablenotes}[flushleft]
						\large
						\item [*]Beginning of the night
						\item \textbf{Notes: }From left to right: the observing night, the number of observed spectra
						(N$_{\mathrm{obs}}$), the exposure time, the average S/N (S/N$_{\mathrm{avg}}$), the S/N variation (S/N$_{\mathrm{min}}$-S/N$_{\mathrm{max}}$) across the whole GIANO-B spectral range, and the airmass.
						
					\end{tablenotes}
			\end{threeparttable}}
		\end{table}
		
		\begin{table}[h]
			\caption{Discarded orders in the \textbf{Basic data reduction} step of the analysis, i.e. alignment in the telluric rest frame and refinement of the wavelength solution.}
			\resizebox{0.99\linewidth}{!}{
				\begin{threeparttable}
					
					\label{bad_orders}
					%\small
					\begin{tabular}{c  l }
						\hline \hline
						\textbf{Date$^*$}  & \textbf{Discarded orders} \\
						\hline
						2019-07-24 & 9, 10, 15, 23, 24, 39, 41, 42, 43, 44, 45, 46, 47, 48, 49\\
						2020-08-09 & 8, 9, 10, 22, 23, 24, 25, 41, 42, 43\\
						2021-10-28 & 8, 9 ,10, 23, 24, 31, 40, 41, 42, 43, 44, 49\\
						\hline
					\end{tabular}
					\begin{tablenotes}
						\large
						\item[*]Start of night date
					\end{tablenotes}
			\end{threeparttable}}
		\end{table}
		
		\begin{table}[h]
			%\centering
			
			\caption{Comparison of atmospheric "full-models". }
			\resizebox{0.9\linewidth}{!}{
				\begin{threeparttable}
					\label{lmax}
						\scriptsize
					\begin{tabular}{c|c c}
					
						\hline \hline
						Model & LH$_{\mathrm{max}}$ & Goodness of fit ($\sigma$)\\
						\hline
						(\textit{i}) & 5527575.71 & 3.38\\
						(\textit{ii}) & 5527581.05 & 0.87 \\
						(\textit{iii}) & 5527581.43 & / \\
						\hline
					\end{tabular}
					\begin{tablenotes}[flushleft]
						\scriptsize
						\item \textbf{Notes: }From left to right: the models name, the maximum of the likelihood matrix LH$_{\mathrm{max}}$, and the goodness of fit obtained with the Wilk's theorem \citep{wilks} on the LH-ratio test. The goodness of fit of the models is shown with respect to the best model in units of standard deviations $\sigma$ (the higher $\sigma$, the more disfavoured the model).
					\end{tablenotes}
			\end{threeparttable}}
			
		\end{table}
		
		\begin{figure}[h]
			\centering
		
			\includegraphics[width=8cm,height=18cm]{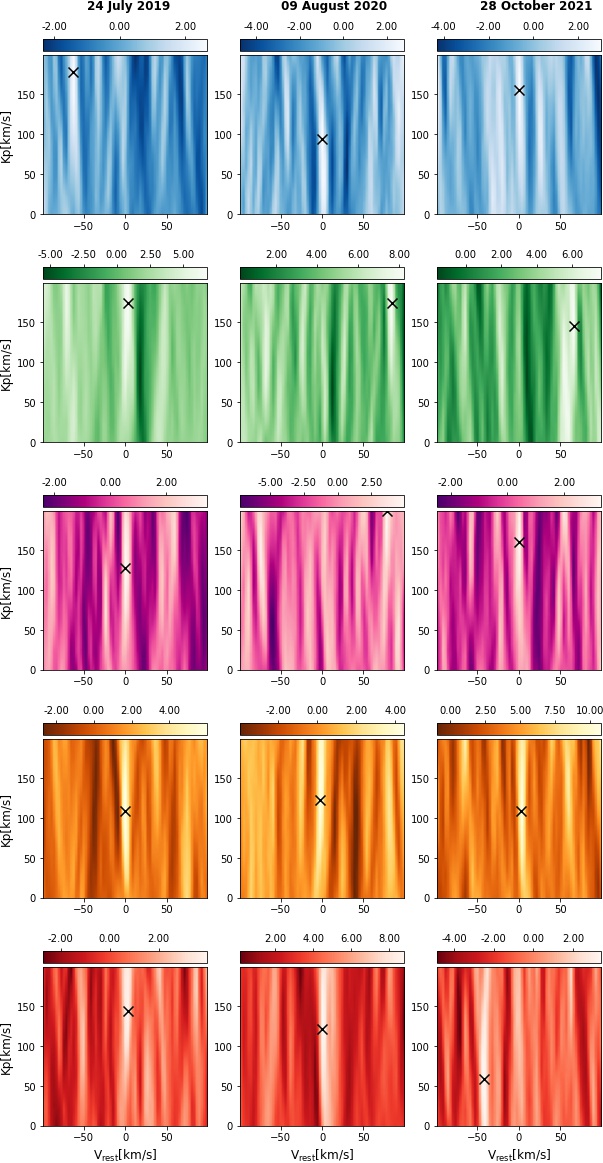}
			\caption{S/N maps for the detected molecules for each night considered individually. The CC peaks are indicated by a black cross. As discussed in \citet{Giacobbe2021}, species have weaker signatures and are not always detected in each individual transit. Nevertheless, they are firmly detected when more transits are co-added. These results show the advantage of our multi-night approach.}
			\label{single_night}
		\end{figure}
		
		\newpage
		
		\section{The \textsc{Pyrat Bay} Code}
		\label{pyratbay}
		
		To compute atmospheric profiles in radiative and thermochemical
		equilibrium, we employed the \textsc{Pyrat Bay} atmospheric modelling
		code \citep{CubillosBlecic2021mnrasPyratBay}.  To evaluate the 
		layer-by-layer fluxes we implemented the two-stream radiative-transfer
		scheme of \citet[][Appendix B]{HengEtal2014apjsTwoStreamRT}.
		This calculation does not consider scattering nor feedback by
		condensates formation.
		
		Following \citet{MalikEtal2017ajHELIOS}, the code attains
		radiative equilibrium by iteratively updating the temperature profile
		such that the divergence of net flux traversing each layer of the
		atmosphere converges to a negligible value.  After each
		temperature update, the \textsc{Pyrat Bay} code re-evaluates the
		atmospheric composition using the thermochemical-equilibrium
		abundances code TEA \citep[][]{BlecicEtal2016apsjTEA}.  This is done
		for a reduced chemical network consisting of H, He, C, N, O, Na, and
		K-bearing species.
		At the top of the atmosphere we imposed an incident stellar
		irradiation from a Kurucz model \citep{CastelliKurucz2003aiausATLAS9}
		according to the physical properties of WASP-69, assuming zero Bond
		albedo and full day--night energy redistribution.
		At the bottom of the atmosphere we imposed an internal radiative
		heat corresponding to a 100~K blackbody.
		
		The atmospheric model spans over a fixed pressure
		range from 100 to $10^{-9}$~bar, and a wavelength grid
		ranging from 0.3 to 30 {$\mu$m} sampled with at a resolving power of $R=15\,000$, sufficient to contain the bulk of the stellar optical
		radiation and planetary infrared radiation.
		The opacity sources include line-list data for the seven molecules
		searched in Section \ref{data_analysis}:
		CO, {\carbdiox}, and {\methane} from HITEMP \citep{COandCO2,
			li_co_2015, CH4_hitemp}; and {\water}, HCN, NH$_3$, and
		C$_2$H$_2$ from ExoMol
		\citep{polyansky_h2o,
			acety,
			YurchenkoEtal2011mnrasNH3opacities,
			HarrisEtal2006mnrasHCNlineList, HarrisEtal2008mnrasExomolHCN,
			nh3}.
		%%%
		Because of the large size of the ExoMol opacities (several billions of
		line transitions), we employed the \textsc{repack} algorithm
		\citep{Cubillos2017apjRepack} to extract only the dominant
		transitions, reducing the number of transitions by a factor of
		$\sim$100 without a significant impact on the resulting opacities.
		%%%
		Additionally, the model included Na and K resonance-line opacities
		\citep{BurrowsEtal2000apjBDspectra}; Rayleigh opacity for
		H, H$_2$, and He \citep{Kurucz1970saorsAtlas}; and collision-induced
		absorption for H$_2$--H$_2$ and H$_2$--He
		\citep{BorysowEtal2001jqsrtH2H2highT,
			Borysow2002jqsrtH2H2lowT, RICHARD2012}.\\
		We considered a few broad regimes by varying two defining properties:
		elemental metallicities sub-solar to super-solar {\em vs.} C/O ratios
		lower and greater than one.  Once we estimated these equilibrium
		models, we post-produced transmission spectra, generated at a high resolving
		power ($R=80\,000$) and then convolved to the GIANO-B instrumental resolution ${\mathbf (R\approx 50\,000)}$, which we
		inspected to determine which species show detectable features over the
		observed wavelength range.

	\end{appendix}

\end{document}